\documentclass[twocolumn,aps,prl,showpacs,groupedaddress]{revtex4}
\pdfoutput=1

\usepackage{amsmath}
\usepackage{dcolumn}
\usepackage{epsfig}
\usepackage{graphicx}
\usepackage{latexsym}

\usepackage{epstopdf}

\begin{document}

\title{Spin gradient thermometry for ultracold atoms in optical lattices}

\author{David M. Weld}
\author{Patrick Medley}
\author{Hirokazu Miyake}
\author{David Hucul}
\author{David E. Pritchard}
\author{Wolfgang Ketterle}
\affiliation{MIT-Harvard Center for Ultracold Atoms, Research Laboratory of Electronics, and Department of Physics, Massachusetts Institute of Technology, Cambridge MA 02139}

\begin{abstract}

We demonstrate spin gradient thermometry, a new general method of measuring the temperature of ultracold atoms in optical lattices.  We realize a mixture of spins separated by a magnetic field gradient.  Measurement of the width of the transition layer between the two spin domains serves as a new method of thermometry which is observed to work over a broad range of lattice depths and temperatures, including in the Mott insulator regime. We demonstrate the thermometry using ultracold rubidium atoms, and suggest that interesting spin physics can be realized in this system.  The lowest measured temperature is 1 nK, indicating that the system has reached the quantum regime, where insulating shells are separated by superfluid layers. 
\end{abstract}


\maketitle

Ultracold atoms trapped in optical lattices represent a new frontier for the investigation of many-body physics \cite{zwerger-review, lewenstein-review}.  The existence of novel physics at decreasing energy scales drives the quest for lower temperatures in the atomic Mott insulator.  Insulating Mott shells form at a temperature $T\sim 0.2U$, where $U$ is the interaction energy.  At the lower temperature $T\sim zJ$, where $J$ is the tunneling amplitude and $z$ is the number of nearest-neighbors, the conducting layers become superfluid and the system enters a quantum insulator state \cite{gerbier-mott}.  At the even colder temperature scale $T\sim J^2/U$, superexchange-stabilized phases can exist in the two-component Mott insulator; this is the regime of quantum magnetism \cite{ho-mott}.  Various proposals~\cite{duandemlerlukin, hofstetterlatticephasediagram} have focused on the realization of quantum spin Hamiltonians in this regime.  Detection of superexchange-driven phase transitions in these systems remains a major goal of ultracold atomic physics.  Perhaps the most important barrier to experimental detection of such a phase transition is the requirement of temperatures well below 1 nK~\cite{ho-mott}.  Additional cooling methods~\cite{cirac-cooling,zoller-cooling,coolingbyshaping,ho-cooling} will be needed to reach this very interesting temperature scale. However, it is clear that to assess current methods and to validate future cooling techniques, low-temperature thermometry of the Mott insulator is needed.

Thermometry of systems in the Mott insulating state has remained a challenge \cite{gerbier-mott, pollet-latticerampup, demarcothermometry, hoffmann-thermometry,capogrosso-thermometry}.   In this paper, we discuss and demonstrate a simple and direct method of thermometry using a magnetic field gradient which works in the two-component Mott insulator.  

The theory behind this method of thermometry is straightforward.  The system under consideration is an ensemble of atoms in a mixture of two hyperfine states loaded into a three-dimensional optical lattice in the presence of a weak magnetic field gradient.  The two states have different magnetic moments, and are thus pulled towards opposite sides of the trapped sample by the gradient.  At zero temperature, the spins will segregate completely, and a sharp domain wall will exist between the two spin domains (a small width due to superexchange coupling is typically negligible).  This system has the same bulk physics as the single-component Mott insulator, but includes additional degrees of freedom in the form of spin excitations in the domain wall.   At finite temperature, spin excitations will increase the width of the domain wall.  This width will depend in a simple way on the field gradient, the differential Zeeman shift, and the temperature, and can thus be used as a thermometer.

For an incoherent mixture of two spins, the partition function for an individual lattice site can be approximately factorized as \mbox{$Z=Z_\sigma Z_0$,} where \mbox{$Z_\sigma= \sum_{\sigma} \exp(-\beta \mbox{\boldmath$\mu$}_\sigma\!\cdot\! \textbf{B}(x))$,} $\beta$ is $1/k_BT$, $\mbox{\boldmath$\mu$}_\sigma$ is the magnetic moment of the spin $\sigma$, $\textbf{B}(x)$ is the spatially varying magnetic field, and $Z_0$ is the partition function of the particle-hole degrees of freedom (for which see \cite{gerbier-mott}).  This approximation is generally valid for the case of one atom per lattice site; for occupation number $n>1$, it is valid when the mean of the intra-spin interaction energies $\overline{U_\sigma}$ is equal to the inter-spin interaction energy $U_{\uparrow\downarrow}$, which is a good approximation in $^{87}$Rb \cite{otheratomsnote}.  Since the total magnetization is fixed, the average value of the magnetic field is cancelled by the corresponding Lagrange multiplier; we include this in the definition of $\textbf{B}(x)$.   We are free to treat the two states as having pseudospin $+1$ and $-1$; making that identification, the mean spin $\langle s \rangle$ as a function of position, gradient strength, and temperature has the simple form
\begin{equation}
\langle s \rangle=\tanh(-\beta\cdot\Delta \mbox{\boldmath$\mu$}\cdot\textbf{B}(x)/2),
\label{tanh}
 \end{equation}
 where $\Delta \mbox{\boldmath$\mu$}$ is the difference between the magnetic moments of the two states.  A fit of the measured spin distribution with a function of this form will give the temperature of the system.  When the Zeeman shift due to the magnetic field gradient is a linear function of position, imaging of the spin distribution essentially corresponds to direct imaging of the Boltzmann distribution.

\begin{figure}[tbh!]\begin{center}
\includegraphics[width=0.8\columnwidth]{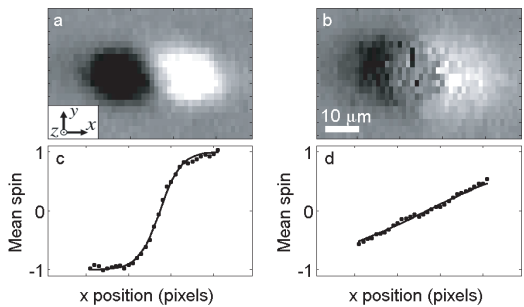}
\caption[fittingexample]{Images used for spin gradient thermometry.  Data on the left were taken at a lower optical trap power than data on the right.  Panels $\textbf{a}$ and $\textbf{b}$ are images of the spin distribution.  Panels $\textbf{c}$ and $\textbf{d}$ show the mean spin versus $x$ position.  The fit to $\textbf{c}$ gives a temperature of 52 nK; the fit to $\textbf{d}$ gives a temperature of 296 nK.  The inset of $\textbf{a}$ shows the axes referred to in the text. The bar in $\textbf{b}$ is a size scale. \label{fittingexample}}
\end{center}\end{figure}

The apparatus used to produce ultracold $^{87}$Rb atoms is described in Ref.~\cite{machinepaper}.  After cooling, approximately $10^5$ atoms are held in a far-red-detuned crossed optical dipole trap with trap frequencies between 100 and 200 Hz.  A three-dimensional cubic optical lattice, formed by three retroreflected beams each of radius $\sim$150$\mu$m, overlaps the trapping region.  Since spin gradient thermometry does not depend on the number of atoms per lattice site $n$, we perform measurements at a range of $n$ values between 1 and 4.  The trapping and lattice beams are all derived from one fiber laser, with a wavelength $\lambda$ of 1064 nm.  Magnetic field gradients up to a few G/cm can be applied with external coils, and calibrated using Stern-Gerlach separation of the different spin states after release from the trap.  The gradient is applied along the $x$ direction, which is the weakest axis of the crossed dipole trap.  Absorptive imaging of the atoms is performed with a camera pointing down along the vertical $z$ axis.

The sequence of steps used to measure temperature is as follows.  First, a sample of $^{87}$Rb atoms in the \mbox{$|F=1,m_F=-1\rangle$} state is prepared by evaporation in the optical trap.  Here $F$ and $m_F$ are the quantum numbers for the total spin and its projection on the $z$ axis, respectively. The atoms are then placed into a mixture of the $|1,-1\rangle$ and $|2,-2\rangle$ states by a nonadiabatic magnetic field sweep through the microwave transition between the two states.  This pair of states was chosen in order to avoid spin-exchange collisions.  A magnetic field gradient of 2 G/cm is applied along the weak axis of the trap and results in additional evaporation, which is intended to remove the entropy created by the state preparation~\cite{chen-gradevap}.  At this point, the field gradient is changed to the value to be used for measurement; lower gradients are used for lower-temperature measurements to keep the domain wall width larger than the imaging resolution.  The optical lattice is then adiabatically ramped up, typically to a depth of 14.5 $E_R$, where $E_R=h^2/2 m \lambda^2$ is the recoil energy and $m$ is the atomic mass.  The transition to the Mott insulator occurs at 13.5 $E_R$. At this point, the spin structure depends on the temperature as discussed above.  

There are several ways to measure the resulting spin distribution.  One way is to first take an image of the $F=2$ atoms in the 14.5 $E_R$ lattice, then in a second run to illuminate the atoms with an optical repumper beam resonant with the $F=1$ to $F'=2$ transition for a few $\mu${}s prior to imaging.  This method gives an image of all atoms and an image of just the $F=2$ atoms; appropriate subtraction can provide the spin distribution.   It is possible to determine the temperature from a single image of one spin, but the data in this paper were all taken using pairs of images to guard against systematic errors.

The temperature can then be measured by fitting the spin distribution to the hyperbolic tangent form.  The resulting thermometer has high dynamic range and variable sensitivity, works at all accessible temperatures of interest, and requires only the simplest fitting procedures. 

\begin{figure}[tbh!]\begin{center}
\includegraphics[width=\columnwidth]{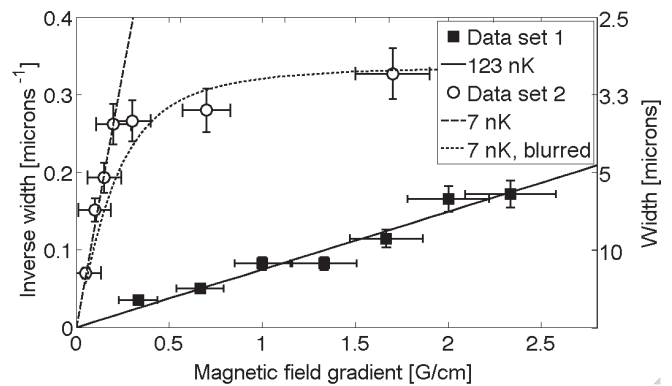}
\caption[widthvsgrad]{Independence of the measured temperature on the applied field gradient.  The inverse of the width of the spin profile is plotted as a function of magnetic field gradient for two data sets at two different temperatures.  For constant temperature, a linear curve is expected.  The width is defined as the distance from the center to the position where the mean spin is $1/2$. The solid (dashed) line assumes a temperature of 123 nK (7 nK) and perfect imaging.  The measured width of the colder data set saturates at high gradient because of finite imaging resolution.  The dotted line assumes a temperature of 7 nK and an imaging resolution of 4 $\mu$m.\label{widthvsgrad}}
\end{center}\end{figure}

Figure~\ref{fittingexample} shows data of the type used for spin gradient thermometry.  An image of the total atom density and an image of the spin density are obtained as discussed above.  Both images are then integrated along the $y$ direction, which is transverse to the gradient.  The spin distribution is then fit by a function of the form $\rho(x) \times \tanh(\frac{3}{4}\beta \mu_B \frac{d |\textbf{B}|}{d x} x)$, where $\rho(x)$ is the total density distribution.  The only free parameters in this fit are a horizontal and vertical offset and the temperature $T=1/k_B\beta$.

\begin{figure}[t]\begin{center}
\includegraphics[width=0.85\columnwidth]{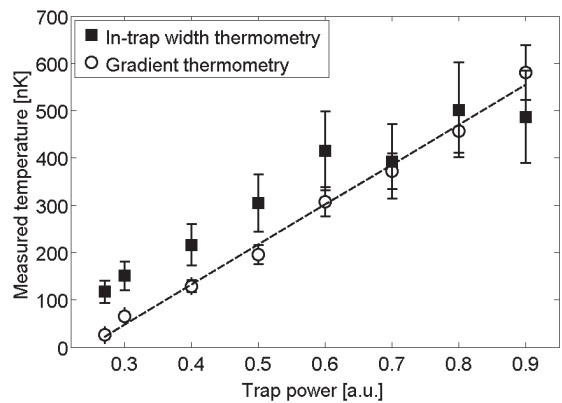}
\caption[tvsodt3]{Validation of spin gradient thermometry.  Comparison of two measured temperatures versus final power in one of the optical trapping beams.  Squares represent the results of in-trap cloud width thermometry, and circles represent the results of spin gradient thermometry (see text for details).  Error bars represent estimated uncertainties.  The dashed line is a linear fit to the spin gradient thermometry data.  The closeness of this fit suggests that the temperature reached is proportional to the trap depth. \label{tvsodt3}}
\end{center}\end{figure}

Figures~\ref{widthvsgrad} and \ref{tvsodt3} show the results of this thermometry on ultracold $^{87}$Rb atoms in an optical lattice.  Figure~\ref{widthvsgrad} shows the linear scaling of the inverse width of the domain wall as the magnetic field gradient is varied while holding the temperature constant.  For widths larger than the optical resolution, the scaling is as predicted by Eq.~\ref{tanh}.  The two data sets plotted in Fig.~\ref{widthvsgrad} were taken at two different temperatures: 7 nK and 123 nK, according to the best-fit theoretical lines.  Finite optical resolution or motion of the atoms during imaging will blur the measured spin profile and result in an overestimate of the domain wall width at high gradients.  This effect was modeled by applying a gaussian blur of radius 4 $\mu$m to the theoretical 7 nK spin profile at various gradients.  The resulting curve, plotted as a dash-dot line in Fig.~\ref{widthvsgrad}, reproduces the saturation of measured width observed in the experimental data.  The effect of finite resolution is always to overestimate the temperature.

Figure~\ref{tvsodt3} shows the measured temperature plotted as a function of the power in the dipole trapping beam which confines the atoms in the direction of the magnetic field gradient (the $x$ direction).  Higher powers in this beam lead to less effective evaporation, and thus higher final temperatures.  As a check of the new method of thermometry, Fig.~\ref{tvsodt3} also presents an analysis of the same data using an existing method of thermometry, based on measurement of the in-trap width of the atomic cloud along the direction perpendicular to the gradient.  This second method is based on the well-known relation $\sigma^2=k_B T /m \omega^2$, where $\sigma$ is the $1/e^2$ half-width of the atomic cloud and $\omega$ is the trap frequency in the direction along which the width is measured \cite{demarcothermometry}.   The width is determined by a fit to the wings of the trapped cloud.  Trap width thermometry is based on a non-interacting approximation, and will fail at temperatures less than $U$ when the system starts to become incompressible.  As in Ref.~\cite{demarcothermometry}, all points on this plot are in the high-temperature single-band regime ($T$ is less than the bandgap but greater than the bandwidth).   For the temperatures plotted in Fig.~\ref{tvsodt3}, the agreement between the two methods is reasonably good, and gives confidence in the use of spin gradient thermometry in regions of parameter space where no other thermometer exists.  

The large dynamic range of spin gradient thermometry is evident in Fig.~\ref{tvsodt3}.  Thermometry can be performed at temperatures so high that no condensate exists before lattice rampup.  The lowest temperature we have measured was achieved by using the new thermometry as a feedback signal, enabling adjustment of experimental parameters for optimization of the final temperature in the Mott insulator. This method allowed us to achieve a measured temperature as low as 1 nK.  At the lattice depth used here, $U$ is 37 nK, and $zJ$ is 6 nK.  The measured temperature is thus well below $T_c=zJ$, the predicted critical temperature for the superfluid layer between the $n=1$ and $n=0$ Mott domains.  According to the treatment of Ref.~\cite{gerbier-mott}, at 1 nK the system should be well inside the quantum regime, with concentric quantum insulator shells separated by superfluid layers.  This represents the first direct demonstration that this temperature regime has been achieved in the Mott insulator.  

At a given value of the magnetic field gradient, very low temperatures will result in a width of the transition region smaller than the imaging optics can resolve (see Fig.~\ref{widthvsgrad}).  However, the width can be increased by decreasing the magnetic field gradient.  The lowest measurable temperature will then depend on the minimum achievable gradient as well as the optical resolution, which are technical rather than fundamental limitations.  In our apparatus, background gradients with all coils turned off are of order $10^{-3}$ G/cm, which, given our imaging resolution of a few $\mu$m, would in principle allow measurement of temperatures down to \mbox{$\sim50$ pK} or the superexchange scale, whichever is higher.

It is instructive to compare the useful range of this new method of thermometry with that of existing methods.  To facilitate meaningful comparison with non-lattice-based methods, we discuss the range of entropy per particle $S/Nk_B$ at which a given thermometer works, rather than the range of temperature.  Condensate fraction thermometry works for $0.35<S/Nk_B<3.5$, where the lower limit is set by the difficulty of detecting a thermal fraction less than 10\%, and the upper limit is set by disappearance of the condensate.  Thermometry based on the thermal cloud size has a similar lower bound, but extends to arbitrarily high values of $S/Nk_B$.  Quantitative thermometry based on the visibility of interference peaks upon release from the lattice requires state-of-the-art QMC calculations fitted to the data.  This technique was recently used to measure temperatures as low as 0.08$U$ in the superfluid phase near the Mott insulator transition \cite{amherst-mainz-peakfitting}.  This method cannot be applied deep in the Mott insulating state \cite{pollet-latticerampup}. Measurement of the width of the conducting layers between the Mott shells is the only previously proposed method which works directly in the Mott insulating state \cite{gerbier-mott, ho-mott, mainztomography}. However, this method requires tomographic techniques, and the useful range of entropy is rather narrow: $0.4<S/Nk_B<\ln(2)$, where the upper limit is set by the melting of the Mott shells, and the lower limit is an estimate based on typical trapping parameters and optical resolution.  Counting only spin excitations, the range of \emph{spin} entropy per particle at which spin gradient thermometry works in our system is $0.1<S_\sigma/Nk_B<\ln(2)$, where the lower limit is a function of optical resolution and sample size and the upper limit corresponds to the point at which the domain wall becomes as wide as the sample.  It is important to note that spin gradient thermometry can work even if the entropy of the particle-hole excitations lies outside of this range in either direction.  For example, spin gradient thermometry can work at arbitrarily high values of the total entropy per particle $S/Nk_B$, assuming the field gradient is increased to the point where $S_\sigma/Nk_B<\ln(2)$.

The method of thermometry presented here works because the two-component Mott insulator in a field gradient has a spectrum of soft and easily measurable spin excitations.  The wide dynamic range of this method is a result of the fact that, in contrast to the gapped spectrum of the bulk one-component Mott insulator, the energy of the spin excitations can be tuned by adjusting the strength of the magnetic field gradient.  The addition of a field gradient and a second spin component does not change the bulk properties of the Mott insulator and can be regarded as ``attaching'' a general thermometer to the first component.

The two component Mott insulator in a field gradient is a rich system which can provide experimental access to novel spin physics as well as thermometry.  In the work presented here, we have always allowed the spin distribution to equilibrate in the gradient before ramping up the optical lattice.  However, changing the gradient after the atoms were already loaded into the lattice should open up several interesting scientific opportunities, in which the gradient is used to manipulate or perturb the atoms rather than as a diagnostic tool.   If, for example, the gradient were suddenly changed after lattice rampup, one could probe non-equilibrium spin dynamics in a many-body quantum system.  If the gradient were instead lowered adiabatically after rampup, adiabatic cooling of the Mott insulator could potentially be performed  which, in contrast to \cite{pfaudemag}, would not involve spin-flip collisions.  

In conclusion, we have proposed and demonstrated a new method of thermometry for ultracold atoms in optical lattices.  We have used the new method to measure temperatures in the Mott insulator as low as 1 nK.  This temperature is to the best of our knowledge the lowest ever measured in a lattice, and it indicates that the system is deep in the quantum Mott regime.

The authors thank Eugene Demler, Takuya Kitagawa, David Pekker, and Lode Pollet for helpful discussions.  This work was supported by the NSF, through a MURI program, and under ARO Grant No. W911NF-07-1-0493 with funds from the DARPA OLE program.  H.M. acknowledges support from the NSF Graduate Research Fellowship Program.

\bibliographystyle{aip}
%\bibliography{/Applications/Bibdesk/dmwbibliography}

\begin{thebibliography}{10}
\bibitem{otheratomsnote}
For other atomic species, this treatment can readily be extended to include
  effects of finite $\Delta U=\overline{U_\sigma}-U_{\uparrow\downarrow}$. The
  leading correction changes the magnetic field gradient at the center of the
  sample $B^{\prime}$ to an effective gradient \mbox{$B^{\prime}(1+(n-1)\Delta
  U/k_B T)$}.
\end{thebibliography}

\end{document}